\documentstyle[sprocl]{article}

\input{psfig}

\def\be{\begin{equation}}
\def\ee{\end{equation}}
\def\bea{\begin{eqnarray}}
\def\eea{\end{eqnarray}}

\def\cmm2{{\,\rm cm^{-2}}}
\def\cm2{{\,{\rm cm}^2}}
\def\cmm3{{\,{\rm cm}^{-3}}}
\def\gcmm3{{\,{\rm g\,cm^{-3}}}}

\def\fun#1#2{\lower3.6pt\vbox{\baselineskip0pt\lineskip.9pt
 \ialign{$\mathsurround=0pt#1\hfil##\hfil$\crcr#2\crcr\sim\crcr}}}

\begin{document}

\title{Primordial Black Hole Formation from Inflaton}

\author{ Xin He MENG$^1$,  Bin  WANG$^2$  
and  S.Feng$^3$}

\address{CCAST B.O.Box 8370, Beijing\\
1.Departments of  Physics, Nankai  University \\
Tianjin, P.R.China\\ Department of Physics, Lancaster University, UK\\
E-mail: mengxh@public1.tpt.tj.cn}

\address{2.Department of Physics,  Shanghai Normal University, P.R.China\\
3.Department of Physics,  University of Science and Technology,
P.R.China\\
}

\maketitle\abstracts{
Measurements of the distances to SNe Ia have produced strong evidence
that
the Universe is really accelarating, implying the existence of a nearly
uniform
component of dark energy with the simplest explanation as a cosmological
constant. In this paper a small changing cosmological term is proposed,
which
is a function of a slow-rolling scalar field, by which the de Sitter
primordial
 black holes' properties, for both charged  and uncharged cases,   
are carefully examined and the relationship between the black hole
formation and the energy
transfer of the inflaton within this cosmological term is
eluciated.}

There is now prima facie evidence that supports
two basic tenets, inflation and dark (matter and energy) components of
the hot big bang universe
paradigm\cite{tu}. Measurements of the distances to SNe Ia have produced
strong evidence that
the Universe is indeed accelarating
which indicates
that most of the critical density exists in the form of nearly uniform and
positive dark energy.
This component is poorly understood. So, naturally
the identification and elucidation of the mysterious
dark-energy component is a very pressing question for nowadays physics.
  Vacuum energy is only the
simplest possibility for the smooth dark component; there are
other possibilities\cite{tur}:  frustrated topological defects
or a slow rolling scalar field or quintessence.   Independent evidence for the existence
of this dark energy, e.g., by CMB anisotropy, the SDSS and 2dF
surveys, or gravitational
lensing, is crucial for verifying the accounting of matter and energy
in the Universe.  Additional and more precious measurements of SNe Ia
could
help shed light on the precise nature of the dark energy.  The dark   
energy problem is not only of great importance for cosmology, but also
for fundamental physics as well.  Whether it is vacuum energy or
quintessence, it is still a puzzle for fundamental physics and possibly
a clue about the unification of the forces and particles\cite{turn}.

For the not very clear dark matter identity, Primordial black holes (PBHs)
are also one of the
possible  cold dark matter candidates
which are believed to take the majority of the matter contents of the
Universe. PBHs  may form in the early universe when pre-existing
adiabatic density fluctuations enter into the cosmological horizon and
recollapse. That is,
primordial overdensities seeded, for instance by inflation, may collapse
to primordial
black holes during early eras if they exceed a critical
threshold\cite{No2}. Thus, it is quite reasonable to discuss PBHs by connecting 
the mysterious dark energy problems to PBHs' formation in the de Sitter spacetimes where a
cosmological term is essential to describe the PBHs properties.

In this paper we propose a tiny changing cosmological term dependent on a
slow-rolling scalar field, which may come from a supersymmetric particle
physics model at higher
energy scale, just as some classes of quintessence 
which may originate from the dynamic supersymmetry breaking 
 of a  supersymmetry particle theory
with a flat direction\cite{ly}. We discuss its
relation
with de-Sitter primordial black holes formation for both charged and uncharged cases, as well as
the black holes's properties.

 Studies of black hole formation from gravitational collapse of a
 (massless)  scalar field have revealed interesting nonperturbative and non-linear energy
(mass) converting  phenomena
 at the threshold of black hole formation\cite{Ch}. Specifically,
 starting from the spherically symmetrical de Sitter black hole spacetimes with charges q
and mass m,
 \be
 ds^2 = - a(t, r)dt^{2}
 + a^{-1}(t, r)dr^{2}
 + r^{2}d\Omega^{2},
 \ee
 where $d\Omega \equiv d\theta^{2} + \sin^{2}\theta d\varphi^{2}$, and
 $\{x^{\mu}\} = \{t, r, \theta, \varphi\}$ are the usual spherical
 coordinates,
\be
a(t,r)=1-2m/r-\Lambda\times r^2/3+q^2/r^2
\ee
The $\Lambda$ is taken the similar form given by Peebles and Vilenkin\cite{pv} with a notation that
now we
only consider a single component field case for simplicity and the
reduced Planck
mass is set $M_p=(8\pi G)^{-1/2}=1$, as well as  $c=\hbar=1$ 
\be
\Lambda(\phi)=b(\phi^4+M^4)
\ee
where the constant energy parameter is assumed to dominate and the self-coupling
constant $b=1\times10^{-14}$
from the condition that present-day large scale structure grows from
quantum fluctuations frozen into $\phi$ during inflation\cite{ly,mp}  as well as the
considerations that
the present density parameter in matter is $\Omega_m\approx0.3$, with
$\Omega_{\phi}=1-\Omega_m\approx 0.7$ in the inflaton\cite{tu,wh,lid}. Besides, we also
require the cosmological term
satisfy  the flatness conditions at the very early Universe evolution 
stage to ensure its slow
changing property, 
\be
\dot\phi\approx-\Lambda(\phi)'/H
\ee
where H is the Hubble parameter which is a function of the inflaton $\phi$ for the very early
Universe evolution period, the overhead dot referring to derivative to time and the prime
indicating derivative with respect to the inflaton $\phi$. The two
flatness conditions are
\be
(\Lambda(\phi)'/\Lambda(\phi))^2/2 =8 \phi^6/{(M^4+\phi^4)^2}<<1
\ee
 and
\be
\Lambda(\phi)''/\Lambda(\phi)=12 \phi^2/(M^4+\phi^4)<<1
\ee
Generally, for a controllable theory it is necessary that the effective potential
only valid at scales lower than Planck energy scale.
 
We can have this quartic term potential from a simple superpotential, Wess-Zumino
model\cite{jb},
\be
 W=c\phi^3
\ee
(where c is a self-coulping constant) with the following U(1) R-symmetry
\be
 \phi\rightarrow exp(i \beta/3)\phi
\ee 
where  $\beta$ is the transformation parameter and 
\be
W\rightarrow exp(i\beta)W,
\ee
plusing a
cosmological constant-like energy  term. Thus, this symmetry forbiddes the other higher order
terms in
$\phi$  and the resultant potential possesses a $Z_4$ symmetry. Generally, if we
require the system having a $Z_2$ symmetry the potential should also include the $\phi^2$
term\cite{mp,ly,di}, that is the mass term. 

As the treatment in literature\cite{pd} we define a parameter $z=r/m$, and another one, the "
charge-mass-ratio "
parameter $\alpha=q/m$ with $m>0$ as well as $r>0$ and
\be
y=3(z^2-2z+\alpha^2)/z^4   
\ee
It is easy to find when a(t,r)=0
\be
\Lambda m^2=y
\ee
Generally equation (2) when a(t,r)=0  possesses four un-degenerate
solutions and equation 
\be
dy/dz=0
\ee   
has two,  among which in our case, at the moment, only the small value one is relevent to
our following analysis,
that is  
\be
0<\alpha^2<9/8
\ee
and 
\be
z_{-}=3/2-(9/4-2\alpha^2)^{1/2}.
\ee
We have the following observations for the charged de Sitter black hole
spacetimes that there shall exist that (we take $\Lambda$  as the value of the potential)

a.  Two horizons provided  $ \Lambda m^2>y(z_{-}) $,

b.  One horizon if  $\Lambda m^2=y(z_{-})$
and

c.  No horizon when  $\Lambda m^2<y(z_{-}) $

The family of parameter, say, $S[q,m,M,b]$, such that for the value
of
$\Lambda $ not less
than certain value $y(z_{-})/m^2$,
black  holes are formed, otherwise  no black holes are formed.

Then, (b), the critical solution is
 {\em universal} with respect to the family of initial data considered,
\be
b(M^4+\phi^4) m^2=y(z_{-})
\ee
and 
\be
y(z_{-})=3\frac{[(3-\sqrt{9-8\alpha^2})^2/4-3+(9-8\alpha^2)^{1/2}+\alpha^2]}
{[3-(9-8\alpha^2)^{1/2}]^4}
\ee
That is, the right hand side  of Eq.(15) is only a function of the "charge-mass-ratio"
\be
b(M^4+\phi^4) m^2 =f(\alpha)
\ee
In the standard scenario  of inflation the inflating expansion lasts about
 a Planck time with  
some 50 e-foldings to solve mainly the original monopole, geometric
flatness and physical
horizon problems\cite{lid}. After
that the inflaton will execute oscillations around the minimum of the
inflation potential, to
convert its stored energy into the to be created physical world during the
reheating period\cite{br}. In the
charged de Sitter black hole case as we discuss the inflaton energy transfered to  form
a black hole
is constrained by the black hole's  charge-mass-ratio $\alpha$
mathematically, which decides the energy transfering rate to form black holes.  Of
cause, the details
of energy transfer mechanism need more physics inputs and theoretical
considerations, especially that where the charges
come from specifically,  besides some proposals such as in PBHs pair
production\cite{hak} or parameter
resonance in preheating era\cite{frd} if the inflaton coupling with other boson or
charged fermion fields. It is similar and straightway to
discuss black holes in de Sitter spacetime without charge,  which will be
discussed following by taking q=0. In this case the parameter set  only consists of three
elements $S[m,M,b]$ and there still shall exist that, with differences from the charged
case,

a. No horizon provided $3\Lambda>1/m$ (the root is negative),

b. One degenerate horizon from three horizons if $3\Lambda=1/m$ with horizon at
$r=1/\Lambda^{1/2}$, that is,  the horizon increases  with energy input, which is obvious in
the simplest
Schwarzchild metric case and the energy transfering in the uncharged case satisfies
\be
b(M^4+\phi^4)m=1/3
\ee

c. Two distinct horizons when $3\Lambda<1/m$ with horizons at $r_1=2cos(\delta)/\Lambda^{1/2}$
and $r_2=2cos(\delta/3+4\pi/3)/\Lambda^{1/2}$ respectively, under condition
$cos(\delta)=-3m\Lambda^{1/2}$.
And the third root of solutions
turns out being negative.

In this simpler case it is clear to see that there still exists the similar energy
transfering relation Eq.(11), but without additional parameter constraint as the charged 
case and the
classical thermodynamics quantities to be calculated are dependent on the slow changing
inflaton, which is  very interesting.  We will detail the tedious computations 
and publish the results elsewhere\cite{men}.

The cosmological term form chosen as the one given by Peeples and Vilenkin is due
to the following two  reasons. One is the connection to the tree level hybrid inflation
model in the very
early Universe era, which is a very promising theory to confront  all
astrophysics
observations as we have known so far\cite{ly,tu,tur,lind}. Another is from the convincing and
consistent results of recent
analyses to the gravitational lensings,
SNe Ia as well as large-scale structure observations\cite{stc} and
theoretical physics considerations with inconsistence predictions from theories \cite{ly1}
that disfavor classes of quintessence models or a simplest cosmological constant
interpretation.\\ \\

Acknowledgements\\

Xin He Meng is indebted  to valuable discussions on this topic  with Laura
Covi, Robert Brandenberger, Ilia Gogoladze, Christopher Kolda,
David Lyth, Leszek Roszkowski, Lewis Ryder, Goran Senjanovic, R.Tung and
Xinmin Zhang  during his stay at
Lancaster University, UK
and ICTP, Italy. He also thanks Profs S.Randjbar-Daemi, Goran Senjanovic
and A.Smirnov for kind
invitation for one month visit at ICTP.
The authors would all like to express their gratitude to Abdus Salam ICTP
for hospitality
extended to them during this work being completed. This work is partly
supported by grants  from National  Education Ministry and Natural Science foundation of
P.R.China\\

References\\

\end{document}